\documentclass[11pt]{article}

\usepackage{bm}
\usepackage{placeins} 
\usepackage{amsmath} 

\usepackage{amssymb}
\usepackage{amsfonts}
\usepackage{multirow}
\usepackage{amsthm}
\usepackage{natbib}
\usepackage{graphicx}

\usepackage[margin=0.75in]{geometry}

\newtheorem{lemma}{Lemma}

\DeclareMathSymbol{\minus}{\mathord}{operators}{"2D}

\begin{document}

\title{Optimization of Survey Weights under a Large Number of Conflicting Constraints}

\author{Matthew R. Williams\hspace{.2cm}\\
    and \\
    Terrance D. Savitsky \\
    U.S. Bureau of Labor Statistics}


\maketitle

\begin{abstract}
In the analysis of survey data, sampling weights are needed for consistent estimation of the population. However, the original inverse probability weights from the survey sample design are typically modified to account for non-response, to increase efficiency by incorporating auxiliary population information, and to reduce the variability in estimates due to extreme weights. It is often the case that no single set of weights can be found which successfully incorporates all of these modifications because together they induce a large number of constraints and restrictions on the feasible solution space. For example, a unique combination of categorical variables may not be present in the sample data, even if the corresponding population level information is available. Additional requirements for weights to fall within specified ranges may also lead to fewer population level adjustments being incorporated. We present a framework and accompanying computational methods to address this issue of constraint achievement or selection within a restricted space that will produce revised weights with reasonable properties. By combining concepts from generalized raking, ridge and lasso regression, benchmarking of small area estimates, augmentation of state-space equations, path algorithms, and data-cloning, this framework simultaneously selects constraints and provides diagnostics suggesting why a fully constrained solution is not possible. Combinatoric operations such as brute force evaluations of all possible combinations of constraints and restrictions are avoided. We demonstrate this framework by applying alternative methods to post-stratification for the National Survey on Drug Use and Health. We also discuss strategies for scaling up to even larger data sets. Computations were performed in R \citep{R17} and code is available from the authors.
\end{abstract}

\noindent%
{\it Keywords:}
benchmarking, calibration, constrained optimization, LASSO, raking, range restrictions
\vfill

\section{Introduction}\label{sec:intro}
Suppose a sample $S$ of $n$ individuals is taken from a population $U$ of size $N$. 
We assign indicators $\delta_i \in \{0,1\}$ to each individual with probability of selection $P(\delta_i = 1) = \pi_i$. Often the probabilities $\pi_i$ are related to the response of interest $y_i$. The balance of information of the observed sample $y_i \in S$ is different from the balance of information in the population $y_i \in U$. We call such a sampling design \emph{informative}. To account for this imbalance, survey weights $d_i = 1/\pi_i$ are used to reduce bias. For example, an unbiased and consistent estimate of the population mean is
$\hat{\mu} = \sum_{S} d_i y_i / \sum_{S} d_i$. Consistent estimation for more general models can be based on the exponentiated likelihood: $L^{\pi}(\theta) = \prod_{S}p(y_i|\theta)^{d_i}$. Use of this pseudo-likelihood provides for consistent estimation of $\theta$ for a broad class of population models \citep{Savitsky16}.

Without weights $\mathbf{d}$, estimates for $\mu$ and $\theta$ are often biased. However, direct use of weights $\mathbf{d}$ may be problematic. For example, not all sampled units may respond. This may lead to bias or at least under-estimation of totals. The use of sampling weights may also be inefficient, leading to larger variances, particularly for sub-population estimates. Some of these issues may be ameliorated if external population information is available (such as totals $\mathbf{t}_2$). Survey weights can then be adjusted to match population level information $\mathbf{d} \rightarrow \mathbf{w}$ such that $\mathbf{A}_2 \mathbf{w} = \mathbf{t}_2$. Finally, weights can be bounded to increase the stability of resulting estimates, $\mathbf{w} \in \{1,M\}$ for some bound $M$.

The focus of this work are insights and tools rooted in optimization, in particular the adjustment of survey sampling weights to meet equality constraints and range restrictions (bounds). However, we first consider the relationship between the optimization perspective presented here and the underlying goals of inference from surveys and statistical models. The current work is complementary as it seeks to provide tools to arrive at reasonable, usable weights, which can either be used directly or as an input to more complex estimation methods \citep[for example, see][]{Chen17}

\subsection{Combining Auxiliary Data with Survey Data} 
\citet{Breidt17} provide a thorough review of model-assisted estimators which combine survey data and auxiliary information to estimate finite population parameters. Several methods can be reformulated as weight adjustments (independent of the survey response); however, response-specific adjustments (or collections of them) are often desirable. In both cases, variable selection may be needed, subsetting the auxiliary variables (e.g. population totals) for exclusion or for possible penalization \citep{Bocci08, Montanari09}. 

Consider a population mean $\bar{y}_{N}$ and vectors of means of population auxiliary variables $\mathbf{\bar{x}}_{N}$ and a sample vector from the population $\mathbf{y}' = (y_1, \ldots, y_n)$ with corresponding sample weight vector $\mathbf{d}$ and matrix of auxiliary variables $\mathbf{X}$. Then a least squares estimator 
$\hat{y}_{ls} = \mathbf{y'd + y'\Omega X (X'\Omega X)}^{\minus 1}(\mathbf{\bar{x}}_{N} - \mathbf{X'd})$ where $\mathbf{\Omega}$ is a scaling matrix.
This estimator can be formulated as a regression model \citep{Isaki82}:
$\hat{y}_{ls} = \bar{y}_{\pi} +  (\mathbf{\bar{x}}_{N} - \mathbf{\bar{x}}_{\pi})' \hat{\boldsymbol{\beta}}_{ls}$ with $\bar{y}_{\pi} = \mathbf{y'd}$ and $\mathbf{\bar{x}}_{\pi} = \mathbf{X'd}$
or as a weight adjusted Horvitz-Thompson estimator \citep{Deville92}:
$\hat{y}_{ls} = \mathbf{y'w}$ with
\[
\mathbf{w} = \mathbf{d }+ \mathbf{\Omega X (X'\Omega X)}^{\minus 1}(\mathbf{\bar{x}}_{N} - \mathbf{X'd})
= \mathbf{d} + \boldsymbol{\eta}
= \mathbf{d} + \mathbf{\Omega X} \boldsymbol{\lambda}.
\]
Here $\boldsymbol{\eta} = \mathbf{\Omega X} \boldsymbol{\lambda}$ is an element-wise offset to the sampling weights and $\boldsymbol{\lambda}$ is the vector of Lagrange multipliers needed to ensure the simultaneous constraints $\mathbf{X'w} = \mathbf{\bar{x}}_{N}$ are met exactly.

While model-assisted estimation focuses on alternatives to the model 
$(\mathbf{\bar{x}}_{N} - \mathbf{\bar{x}}_{\pi})' \hat{\boldsymbol{\beta}}_{ls}$ and inferences based on 
$\hat{\boldsymbol{\beta}}_{ls} = \mathbf{y'\Omega X (X'\Omega X)}^{\minus 1}$
or difference estimators linked to $\bar{y}_{\pi}$ \citep{Breidt17}, an optimization perspective will focus on the Lagrange multipliers $\boldsymbol{\lambda} = \mathbf{(X'\Omega X)}^{\minus 1}(\mathbf{\bar{x}}_{N} - \mathbf{X'd})$
and offset $\boldsymbol{\eta}$ to address the complexities of constraints and restrictions on $\mathbf{w}$ and strategies for computation. 
For example, \cite{Mcconville17} use a LASSO technique to shrink the $\hat{\boldsymbol{\beta}}_{ls}$ to $\mathbf{0}$, which leads to the \emph{exclusion} of variables (columns of $\mathbf{X}$) and is dependent on the response $\mathbf{y}$. We describe a method that shrinks the discrepancy $(\mathbf{\bar{x}}_{N} - \mathbf{X'd})$ to $\mathbf{0}$ which leads to the \emph{inclusion} of variables and is independent of $\mathbf{y}$.

As an alternative to model-assisted estimation, a predictive modeling approach may be used to motivate estimation of population quantities \citep{Little93}. Like model-assisted estimation, performance of methods and the need for weights in analyses are typically dependent on the response $\mathbf{y}$. Weight adjustment methods that are applied to all responses will lead to varied results, increasing the variance of estimates unrelated to the auxiliary information and decreasing the mean squared errors of estimates related to that information \citep{Little05}. \citet{Salgado12} advocate for loosening the multipurpose requirement (only one weight per observation, regardless of inference) but note several challenges to consider for official statistics.

\subsection{Modelling Concepts}
Outside of survey sampling, the methods presented in this work have connections to penalized likelihood and Bayesian formulations, which we briefly outline here. The linear calibrated weights
$\mathbf{w} = \mathbf{d }+ \mathbf{\Omega X (X'\Omega X)}^{\minus 1}(\mathbf{t} - \mathbf{X'd})$ 
are the solution to a minimization
$\lambda(\mathbf{w}) = (\mathbf{w} - \mathbf{d})' \mathbf{\Omega}^{\minus 1}(\mathbf{w} - \mathbf{d})$
within the constrained space $\mathbf{X'w} = \mathbf{t}$.
If we assume the very simple model for $\mathbf{z} = \mathbf{X}'\mathbf{w}$ with $\mathbf{z} \sim N(\mathbf{t}, \alpha \mathbf{\Sigma})$
and prior distribution $\mathbf{w} \propto N(\mathbf{d}, \mathbf{\Omega})$, then an estimate for $\mathbf{w}$ based on maximizing the posterior would be
\[
\mathbf{w}_{MAP} = \mathbf{d }+ \mathbf{\Omega X} (\mathbf{X'\Omega X} + \alpha\mathbf{\Sigma})^{\minus 1}(\mathbf{t} - \mathbf{X'd})
\]
Scaling the matrix $\mathbf{\Sigma}^{\minus 1}$ (or exponentiating the likelihood) by a factor $\alpha \in (1,\infty)$ is analogous to observing more data and thus increasing the importance of the likelihood relative to the prior. Since direct measurement of individual $w_i$ never occurs in the likelihood, the posterior distribution $p(\mathbf{w}|\mathbf{z})$ converges to a distribution on a simplex defined by $\mathbf{X}'\mathbf{w} = \mathbf{t}$ rather than to a degenerate point distribution. This is the basis of the method of data cloning \citep{Lele07} to evaluate which parameters are identifiable for maximum likelihood estimation by having distributions collapse to points where unidentifiable parameters collapse  to non-degenerate distributions instead of points.
Alternatively, one could also propose that the distribution for $\mathbf{w}$ is the likelihood and the distribution for $\mathbf{z}$ is actually a penalty.
Then $\mathbf{w}_{MAP}$ is also a ridge estimator \citep{Rao97}. \citet{Tibshirani11} consider the case when $\mathbf{z}$ has a Laplace distribution centered at $\mathbf{t}=\mathbf{0}$ and formulate a generalized LASSO.

While methods exist for inference on compositional data \citep[see for example][]{Aitchison82} and benchmarked time series \citep{Pizzinga10}, results are based on linear projections and simple transformations. The optimization perspective in this current work supplements this by considering more complicated support $\mathbf{w} \in \mathbb{C} \subset \mathbb{R}^n$ which involves not only equality constraints, but also inequalities in the form of allowable ranges. While the working models (or objective functions) may be quite simple, the support or `feasible regions' are complex. In particular, this work explores what can be done when these regions are empty (i.e. no solution exists). Application of more sophisticated models could only follow after these complex regions are explored and understood via simpler working models.

\subsection{An Optimization Perspective}
For simplicity of presentation, we focus on survey weight adjustment and refer to it broadly as calibration. Calibration includes post-stratification and extreme-weight adjustment and may include certain types of non-response adjustments \citep{Sarndal05}. 
A major difficulty in calibration could be referred to as constraint identifiability and compatibility with other restrictions. A collection of constraints may be internally inconsistent if the column rank of the predictor matrix is less than the number of predictors. For example, the sample data may not contain observations for every crossing of demographic control variables, such as Age by Race by State. 

When including a restricted solution space such as $\mathbf{w} \in [\mathbf{l}, \mathbf{u}]$, even consistent constraints may lead to an empty feasible region in which no set of points can satisfy all constraints. As an example, consider the calibration of a set of sampling weights to meet a large number of targets such as demographic counts crossed with geographic variables.
Final weights are restricted to fall within a specified range or change a certain percentage from the original sample weights to reduce the impact of extreme values and the increase of variance due to unequal distribution of weights. The result of these additional restrictions is that the feasible region, which would otherwise contain weights satisfying all these restrictions and constraints, is likely to be empty. One approach is to ignore this issue and provide algorithms for achieving a solution when one exists \citep{Singh96}. Variations of relaxing constraints have been developed \citep{Rao97, Therberge00, Fetter05, Montanari09, Bocci08, Park09, Breidt17}. Another approach would be to relax the restrictions (weight ranges) or prioritize targets. This prioritization is then used to iteratively include targets or aggregate (collapse) variable levels \citep{Lamas15, MRB:PersonWeight:2014}.

Through a combination of insights from survey analysis and other areas of statistics such as path-algorithms \citep{Tibshirani11, Zhou13} and data-cloning \citep{Lele07, Lele10}, we present a new framework and accompanying tools for addressing the challenge of simultaneous optimization and constraint selection. The focus shifts from finding one optimal solution to finding one or more reasonable solutions which satisfy most of the constraints and restrictions. This framework provides the flexibility to compromise in a systematic but customizable way including the
\begin{itemize}
	\item Use of the most popular deviance (distance) measures
	\item Use of range restrictions for weights 
	\item Use of point and interval constraints (controls)
	\item Ability to identify and target subsets of weights  and constraints which may be driving non-existence of a solution
\end{itemize}
Results are produced without combinatoric operations such as brute force evaluations of all possible combinations of inclusion/exclusion of constraints and restrictions. Instead, the user is provided with both a partially constrained solution and diagnostics suggesting why a fully constrained solution is not possible.

In Section \ref{sec:motiv}, we establish notation and examine a post-stratification example from the National Survey on Drug Use and Health (NSDUH). Section \ref{sec:approach} describes the general approach as a synthesis of conceptual and computational tools. Section \ref{sec:app} revisits the NSDUH example in more detail. Section \ref{sec:guide} discusses experiences and suggestions for working with larger data sets. Section \ref{sec:ext} provides ideas for future work and points out areas where theory needs to be further developed. Section \ref{sec:conc} reiterates the conclusions for this work. Code for the implementation was created in R \citep{R17} and is available from the authors.

\section{Motivating Example: Post-stratification for the National Survey on Drug Use and Health}\label{sec:motiv}
The following provides baseline notation for optimization which can be modified according to whether our purpose is sample weight calibration:
\begin{itemize}
\item 	Let $\mathbf{y}$ represent the vector of original, unconstrained values ($\mathbf{y}=\mathbf{d}$ for sample weights). We wish to find the closest values $\mathbf{x}$ to $\mathbf{y}$ such that the linear constraints $\mathbf{A}_{1} \mathbf{x}=\mathbf{t}_{1}$ are satisfied (for non-linear constraints, see Section \ref{sec:ext}). We measure `closeness' by a deviance $\delta_{1} (\mathbf{x}|\mathbf{y},\mathbf{Q}_{1})$ with scaling matrix $\mathbf{Q}_{1}$. The simplest example is the quadratic deviance $(\mathbf{x}-\mathbf{y})'\mathbf{Q}_{1} (\mathbf{x}-\mathbf{y})$, but there are several other reasonable choices (see Section \ref{sec:approach}).
\item 	Additionally, we may impose a penalty to pull the linear combination $\mathbf{A}_{2} \mathbf{x}$ towards a `soft' target $\mathbf{t}_2$ which does not have to be met exactly. The penalty can conveniently take the form of $\alpha \delta_{2} (\mathbf{A}_{2} \mathbf{x}|\mathbf{t}_{2},\mathbf{Q}_{2})$, which is a deviance scaled by a penalty factor $\alpha >0$.
\item 	Finally, we impose restrictions on the solution $\mathbf{x}$. For example, we want $\mathbf{b}_{l} \le \mathbf{x} \le \mathbf{b}_{u}$ or more generally $\mathbf{R} \mathbf{x} \le \mathbf{b}$, where $\mathbf{b} = (\mathbf{b}_{l},\mathbf{b}_{u})$ are the lower and upper bounds and $\mathbf{R}$ is a matrix of indicators. This is common for calibration where final weights are often required to be at least 1 and less than some threshold to control for extreme values. 
\end{itemize}

As an example, we examine the final weight adjustment step for the 2014 National Survey on Drug Use and Health (NSDUH), sponsored by the Substance Abuse and Mental Health Services Administration (SAMHSA). NSDUH is the primary source for statistical information on illicit drug use, alcohol use, substance use disorders (SUDs), mental health issues, and their co-occurrence for the civilian, non-institutionalized population of the United States. The final analysis weight is a product of five inverse probabilities of selection corresponding to the five stages of nested sampling (three geographic, household, individual) and  weighting adjustments for non-response, post-stratification, and extreme weight trimming at different stages \citep{MRB:PersonWeight:2014}. The final weighting model is a post-stratification to population control totals and is performed separately for the nine US Census divisions. 

For illustration, we focus on one model for the New England division. Table \ref{tab:analwt} summarizes the demographic variables used for the adjustment including the number of levels proposed and those achieved. The comments for each variable set indicate all the deviations from the proposed variables due to convergence problems (i.e. no solution) and enforcing a hierarchy of constraints (e.g. two-factor levels only included for corresponding one-factor levels included). The process of performing this adjustment was iterative, involving several adjustments when convergence or singularity issues were encountered. Out of 267 possible population controls, 196 constraints were used. If we look at the rank of the observed variables in the sample, the $\mathbf{A}$ matrix is 267 by 5791 and has an estimated rank of 257. Thus we expect to miss at least 10 constraints. Four predictors are not observed in the sample data. Then there are essentially 6 more variables than needed in the sample data. However, reducing the dimension any further is not straight-forward. One reason why the number of achieved targets is lower than 257 is because each weight was restricted to lay within a range of values, bounded away from 0. This additional restriction reduces the number of possible constraints. After discussing the methods and general principles, we will revisit this example to demonstrate that more constraints can be achieved while still maintaining the range restrictions for the weights. Furthermore, this process can be performed in a minimally supervised way without stopping and restarting to drop or collapse variables.

\begin{table}
\begin{center}
\caption{Covariates for 2014 NSDUH Person Weights (res.per.ps), Model Group 1: New England \citep[see][Exhibit D1.5]{MRB:PersonWeight:2014}.
\emph{Coll.} Levels collapsed together; \emph{ Conv.} Model not convergent; \emph{Drop.} Levels collapsed into reference set; \emph{ Sing./zero.} Levels removed due to singularities or zero sample size; \emph{ Keep.} Levels kept and the remainder dropped.}
{\scriptsize
\begin{tabular}{|lrrrl|}
  \hline
{\bf Variables}&{\bf Levels}&{\bf Proposed}&{\bf Final}&{\bf Comments}\\ \hline
\emph{One-Factor Effects}&&\emph{20}&\emph{20}&\\ \hline
Intercept&1&1&1&All levels present.\\
State&6&5&5&All levels present.\\
Quarter&4&3&3&All levels present.\\
Age&6&5&5&All levels present.\\
Race (5 levels)&5&4&4&All levels present.\\
Gender&2&1&1&All levels present.\\
Hispanicity&2&1&1&All levels present.\\ \hline
\emph{Two-Factor Effects}&&\emph{95}&\emph{90}&\\ \hline
Age  x  Race (3 levels)&6  x  3&10&10&All levels present.\\
Age  x  Hispanicity&6  x  2&5&5&All levels present.\\
Age  x  Gender&6  x  2&5&5&All levels present.\\
Race (3 levels)  x  Hispanicity&3  x  2&2&2&All levels present.\\
Race (3 levels)  x  Gender&3  x  2&2&2&All levels present.\\
Hispanicity  x  Gender&2  x  2&1&1&All levels present.\\
State  x  Quarter&6  x  4&15&15&All levels present.\\
State  x  Age&6  x  6&25&25&All levels present.\\
State  x  Race (5 levels)&6  x  5&20&15& Coll. (2,3), (2,4) \& (2,5), repeat\\
	& & & &  VT, coll. (4.3) \& (4,4); conv. \\
State  x  Hispanicity&6  x  2&5&5&All levels present.\\
State  x  Gender&6  x  2&5&5&All levels present.\\ \hline
\emph{Three-Factor Effects}&&\emph{152}&\emph{86}&\\ \hline
Age  x  Race (3 levels)  x  Hispanicity&6  x  3  x  2&10&8& Drop (5,2/3,1); conv.\\
Age  x  Race (3 levels)  x  Gender&6  x  3  x  2&10&8& Drop (5,2/3,1); conv.\\
Age  x  Hispanicity  x  Gender&6  x  2  x  2&5&5&All levels present.\\
Race (3 levels)  x  Hispanicity  x  Gender&3  x  2  x  2&2&2&All levels present.\\
State  x  Age  x  Race (3 levels)&6  x  5  x  3&50&12& Coll. (1,1,2) \& (1,1,3), repeat age\\
 & & & &levels 2, 3, and 4, coll. (2,1,2) \&\\
 & & & &  (2,1,3), coll. (3,1,2) \& (3,1,3), repeat \\
 & & & & age levels 2 and 3, coll. (4,1,2) \& \\
 & & & & (4,1,3), repeat age levels 2 and 3, \\
 & & & & coll. (5,1,2) \& (5,2,3); sing./zero/conv.\\
State  x  Age  x  Hispanicity&6  x  6  x  2&25&12& Drop (1,5,1), (2,*,1), (3,4/5,1), \\
 & & & & (4,5,1), (5,2/3/4/5,1); \\
 & & & & sing./zero/conv.\\
State  x  Age  x  Gender&6  x  6  x  2&25&25&All levels present.\\
State  x  Race (3 levels)  x  Hispanicity&6  x  3  x  2&10&4& Coll. (1,2,1) \& (1,3,1),  coll. (3,2,1) \& \\
 & & & &(3,3,1), keep (4,2,1), (4,3,1), \\
 & & & &drop others; conv.\\
State  x  Race (3 levels)  x  Gender&6  x  3  x  2&10&6& Drop (2,2/3,1), (5,2/3,1); conv.\\
State  x  Hispanicity  x  Gender&6  x  2  x  2&5&4& Drop (5,1,1); conv.\\ \hline
\emph{Total}&&\emph{267}&\emph{196}&\\ \hline
\end{tabular}
\label{tab:analwt}
}
\end{center}
\end{table}

\FloatBarrier

\section{General Approach}\label{sec:approach}
This section breaks the framework into individual components. We first discuss the conceptual formulation and the versatility of the deviance measure (Section \ref{sec:dev}). We then discuss computational strategies for solving a simplified optimization without penalties (Section \ref{sec:solver}). We next demonstrate how to extend this functionality to include penalization (Section \ref{sec:penalty}). Lastly, we combine all components (Section \ref{sec:synth}) to create a general strategy demonstrated in more detail for the NSDUH example (Section \ref{sec:app}).

\subsection{Choices for Deviances}\label{sec:dev}
First, consider penalization and equality constraints without the additional restrictions on ranges or rounding. The objective function would take the form:
\begin{equation}
\label{eq:optim}
\min_{\mathbf{x}_1} \left\{ \delta_1(\mathbf{x}_1|\mathbf{y},\mathbf{Q}_1) + \alpha \delta_2(\mathbf{A}_2 \mathbf{x}_1|\mathbf{t}_2, \mathbf{Q}_2) \right\},  
\text{ s.t. }  \mathbf{A}_1 \mathbf{x}_1 = \mathbf{t}_1.
\end{equation}

We  can make the following connections:
\begin{itemize}
	\item $\alpha \delta_2$ can be seen as a deviance on the transformed variables $\mathbf{x}_2 = \mathbf{A}_2 \mathbf{x}_1$, with scaling matrix $\alpha \mathbf{Q}_2$.
	\item The equality constraint $\mathbf{A}_1 \mathbf{x}_1 = \mathbf{t}_1$ is equivalent to having another penalty $\beta \delta_3(\mathbf{A}_1 \mathbf{x}_1|\mathbf{t}_1, \mathbf{Q}_3)$  with $\beta \rightarrow \infty$
\end{itemize}
Thus each of the three pieces in the optimization above is related to a deviance $\delta$.

We highlight the quadratic and three other popular choices of smooth deviances and describe the critical non-smooth absolute difference deviance and its extensions. The smooth deviances are differentiable at 0 and can all be considered as members of a general class corresponding to adjustments which take the unconstrained value $\mathbf{y}$ and apply element-wise adjustments $\boldsymbol{\eta}(\mathbf{A,t})$ to obtain the constrained $\mathbf{x} = h[\boldsymbol{\eta}]$ where $[.]$ signifies element-wise operations inside. \citet{Deville92} formulate the adjustment as multiplicative: $h[\boldsymbol{\eta}] = [\mathbf{y} g[\boldsymbol{\eta}]]$. However this is awkward for some deviances, such the quadratic whose adjustment is clearly additive: $h[\boldsymbol{\eta}] = \mathbf{y} + \boldsymbol{\eta}$ vs. $g[\boldsymbol{\eta}] = 1 + [\boldsymbol{\eta}/\mathbf{y}]$. Instead, we generalize slightly by removing this multiplicative formulation and working with $h[\boldsymbol{\eta}]$ directly instead of $g[\boldsymbol{\eta}]$. The non-smooth deviances are critical for regression coefficient and constraint selection \citep{Zhou13, Nocedal06}. They are not differentiable at 0, making computation and analytical properties more complex. However, they can be expressed as the limit of iteratively rescaling members of the smooth deviance class (See Section \ref{sec:penalty}).

\subsubsection{Smooth Deviances}\label{sec:smoothdev}
The four most popular smooth deviances are (i) the quadratic deviance, (ii) the Poisson deviance, (iii) the discrimination information deviance, and (iv) the logistic deviance. 

The quadratic deviance has the simplest form $\delta(\mathbf{x|y,Q}) = \mathbf{(x-y)'Q(x-y)}$ with symmetric, invertible, and usually positive definite or semi-definite scaling matrix $\mathbf{Q}$. Final solutions $\mathbf{x}$ can be expressed as the element-wise additive adjustment $\mathbf{x} = \mathbf{y} + \boldsymbol{\eta}$ with $\mathbf{x,y} \in \mathbb{R}^{n}$. The quadratic deviance is used to obtain the restricted least squares estimator \citep{Greene00}, the generalized regression estimator \citep{Isaki82}, and the ridge estimator \citep{Rao97} when used as a penalty.

The Poisson deviance corresponds to the pseudo empirical likelihood \citep{Chen99, Chen02} with $\delta(\mathbf{x|y,Q}) = \mathbf{1'\left< q\right>\left[y \log[y/x] - y + x\right]}$ where $\left< \mathbf{q}\right>$   is a diagonal matrix, usually positive definite and full rank. The solution $\mathbf{x}$ is a product of the unconstrained $\mathbf{y}$ and a multiplicative adjustment $\mathbf{x} = [\mathbf{y}/(\mathbf{1}-\boldsymbol{\eta})]$ with $\mathbf{x,y} \in \mathbb{R}_{+}^{n}$. This is sometimes referred to as the Likelihood approach.

The discrimination information deviance \citep{Ireland68} is also known as the Kullback-Leibler divergence and is sometimes referred to as exponential tilting \citep{Kim10}. It is the measure being minimized by the procedure known as iterative proportional fitting or `raking' \citep{Deming40}. Raking has many applications, including the adjustment of vector-valued cells \citep{Peitzmeier88}. This deviance measure is very similar to the Poisson deviance with the roles of $\mathbf{x}$ and $\mathbf{y}$ switched: $\delta(\mathbf{x|y,Q}) = \mathbf{1'\left< q\right>\left[x \log[x/y] - x + y\right]}$. This leads to an appealing element-wise multiplicative adjustment $\mathbf{x} = [\mathbf{y}\exp[\boldsymbol{\eta}]]$ with $\mathbf{x,y} \in \mathbb{R}_{+}^{n}$. Combining the Poisson and the discrimination deviances, a symmetric deviance is achieved: $\delta(\mathbf{x|y,Q}) = \mathbf{1'\left< q\right>\left[(x-y) \log[x/y] \right]}$.

The logistic deviance was presented by \citet{Deville92} as one approach to keep weights within specified bounds:
\[
\delta(\mathbf{x|y,Q},\mathbf{b}_l,\mathbf{b}_u) = \mathbf{1'\left< q\right>} \left[(\mathbf{x}-\mathbf{b}_l) \log[(\mathbf{x}-\mathbf{b}_l)/(\mathbf{y}-\mathbf{b}_l)] + (\mathbf{b}_u - \mathbf{x}) \log[(\mathbf{b}_u - \mathbf{x})/(\mathbf{b}_u - \mathbf{y})] \right]
\] 
with limits $\mathbf{x,y} \in [\mathbf{b}_l, \mathbf{b}_u]$. Often, the logistic deviance is parameterized in terms of a ratio adjustment to $\mathbf{y}$ with bounds $\mathbf{b}_{l}^{*} \le \mathbf{1}  \le  \mathbf{b}_{u}^{*}$ \citep{Deville92, Folsom00}. Starting from $\delta(\mathbf{x|y,Q},\mathbf{b}_l,\mathbf{b}_u)$, the logistic deviance could be motivated as a combination of shifted and reflected discrimination information deviances. However its name and purpose are more obvious when viewing the relationship between the final $\mathbf{x}$ and the adjustments to $\mathbf{y}$: $\mathbf{x} = \Phi_{log}\left[\boldsymbol{\eta} + \Phi^{\minus 1}_{log}\left[(\mathbf{y} - \mathbf{b}_l)/(\mathbf{b}_u - \mathbf{b}_l) \right] \right](\mathbf{b}_u - \mathbf{b}_l) + \mathbf{b}_l$ where $\Phi_{log}$ is the probability function for a logistic random variable, $\Phi^{\minus 1}_{log}$ is the quantile function, and $\phi_{log}$  is the density function. 
The solution $\mathbf{x}$ is a scaled, shifted, and re-centered probability function.  Evaluating the logistic distribution functions:
\[
\mathbf{x}=\left(\frac{(\mathbf{y}-\mathbf{b}_l ) e^{\boldsymbol{\eta}}}{(\mathbf{b}_u-\mathbf{y})+(\mathbf{y}-\mathbf{b}_l )} e^{\boldsymbol{\eta}} \right)(\mathbf{b}_u-\mathbf{b}_l )+\mathbf{b}_l=   \frac{\mathbf{b}_l  (\mathbf{b}_u-\mathbf{y})+\mathbf{b}_u (\mathbf{y}-\mathbf{b}_l ) e^{\boldsymbol{\eta}}}{(\mathbf{b}_u-\mathbf{y})+(\mathbf{y}-\mathbf{b}_l ) e^{\boldsymbol{\eta}}}
\]
The formulation on the right \citep{Deville92} tends to obscure the connection to distribution functions. However, retaining the connection has benefits: (i) It suggests trying alternative symmetric distributions such as Gaussian, Laplace, student’s t, etc. Asymmetric distributions might also be used. This would be analogous to alternatives to logistic regression such as probit regression. In practice, there may be little difference in results, but some distributions might lead to computational efficiencies or increased stability. (ii) Using distribution functions directly in computation is more stable, because they are defined on the extended real numbers. For example, the expression $e^{\eta}/(1+e^{\eta})$ yields `not a number' results for software when $\eta = \infty$ and unstable results when $\eta$ is very large. In contrast, statistical software correctly maps $\Phi_{log} (-\infty)= 0$ and $\Phi_{log} (\infty)= 1$ and tolerates large finite inputs. The same is true for density functions. Therefore some scaling and centering tricks needed to stabilize the computation using the formulation of \citet{Deville92} become unnecessary when using the distributional formulation.

\subsubsection{Non-smooth Deviances}\label{sec:absdev}
We now consider non-smooth deviances based on the absolute difference function. The absolute difference measure and its variants are used for quantile estimation \citep{Koenker01} and controlled rounding problems \citep{Cox82}. It is particularly useful as a non-smooth penalty in LASSO regression \citep{Tibshirani96} and constrained optimization \citep{Nocedal06}, because non-smooth functions can achieve subset selection (such as forcing regression coefficients to zero) for finite penalties $(\alpha)$. The deviance $\delta(\mathbf{x|y,Q}) = \mathbf{1'\left< q\right>|x-y|}$ is simple, but the expression for $\mathbf{x}$ is more complicated. Fortunately, we can solve for $\mathbf{x}$ by iteratively rescaling a quadratic deviance \citep{Fan01, Lange00, Hunter04}. Then $\mathbf{x}^k=\mathbf{y}+\boldsymbol{\eta}(\mathbf{x}^{k-1})$, for iterations $k=1,2,\ldots$

\begin{figure}
\centering
\includegraphics[width = 0.45\textwidth,
		page = 1,clip = true, trim = 0in 0.25in 0in 0.25in]{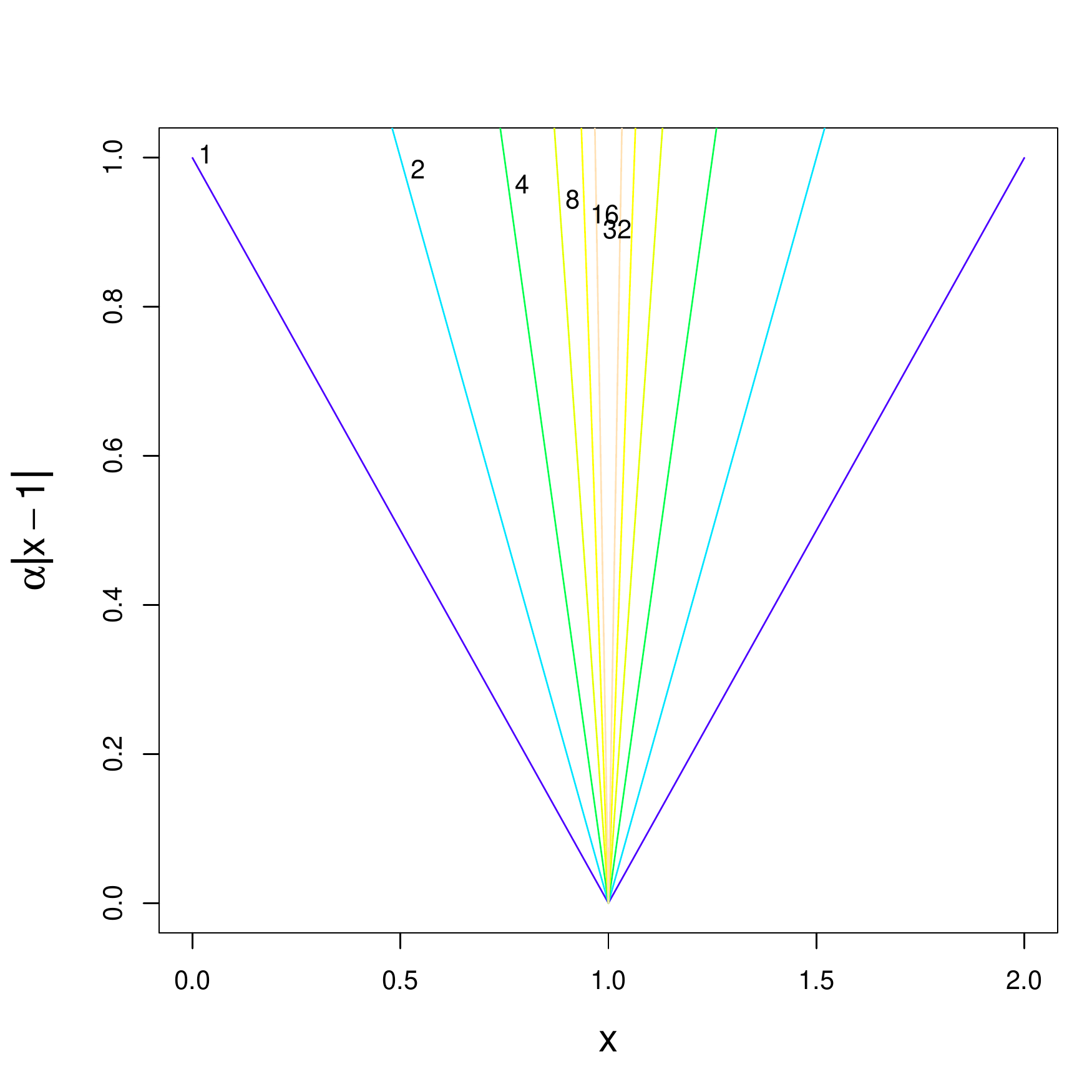}
\includegraphics[width = 0.45\textwidth,
		page = 3,clip = true, trim = 0in 0.25in 0in 0.25in]{Interval_Penalties.pdf}\\
\includegraphics[width = 0.45\textwidth,
		page = 2,clip = true, trim = 0in 0.25in 0in 0.25in]{Interval_Penalties.pdf}
\includegraphics[width = 0.45\textwidth,
		page = 4,clip = true, trim = 0in 0.25in 0in 0.25in]{Interval_Penalties.pdf}
\caption{Non-smooth absolute difference (left) and smooth quadratic (right) penalties. One-piece centered a $x=1$ (top) and two-piece centered at $x = 0.5, 1.5$ (bottom) for 
$\alpha = 1,2,4,8,16,32$ (blue to yellow).}
\label{fig:int_pen}
\end{figure}

It can happen that the input $\mathbf{y}$ is specified as an interval $[\mathbf{y}_l,\mathbf{y}_u]$ rather than a point. The most useful case occurs when we use the penalty deviance $\delta_2(\mathbf{A}_2 \mathbf{x}_1|\mathbf{t}_2, \mathbf{Q}_2)$ from system \ref{eq:optim}. For example, we may want to loosen the bounds on the total number of people reporting two or more race categories for a state by specifying an acceptable range rather than a specific target: $\mathbf{t}_2 \in [\mathbf{t}_l,\mathbf{t}_u]$. 
Then any final weights $\mathbf{x}_1$ which leads to an estimate $\mathbf{A}_2 \mathbf{x}_1 \in [\mathbf{t}_l,\mathbf{t}_u]$ will be equally acceptable by having the same minimum value for the penalty function $\delta_2(\mathbf{A}_2 \mathbf{x}_1|\mathbf{t}_2, \mathbf{Q}_2)$. In order to implement this, we can extend the absolute difference measure centered on the point $\mathbf{y}$ to a measure centered at intervals: 
$\delta(\mathbf{x}|\mathbf{y}_l,\mathbf{y}_u,\mathbf{Q}) = \mathbf{1'\left< q\right>} \left[|\mathbf{x}-\mathbf{y}_l | + |\mathbf{x}-\mathbf{y}_u | - |\mathbf{y}_{u}  - \mathbf{y}_{l} | \right]$. The last term is fixed with respect to $\mathbf{x}$, so we can equivalently use:
\[
\delta(\mathbf{x}|\mathbf{y}_l,\mathbf{y}_u,\mathbf{Q}) = \mathbf{1'\left< q\right>}[|\mathbf{x}-\mathbf{y}_l | + |\mathbf{x}-\mathbf{y}_u |] =
\delta(\mathbf{x}|\mathbf{y}_l,\mathbf{Q}) + \delta(\mathbf{x}|\mathbf{y}_u,\mathbf{Q}),
\]
which is equal to the sum of two absolute difference deviances centered at points $\mathbf{y}_l$ and $\mathbf{y}_u$ with common scaling $\left< \mathbf{q} \right>$. As $\alpha$ increases, the penalty outside $[\mathbf{y}_l,\mathbf{y}_u]$ becomes steeper, forcing values closer to the interval (Figure \ref{fig:int_pen}). In contrast, using a two-piece quadratic penalty at $\mathbf{y}_l$ and $\mathbf{y}_u$ will not achieve the desired effect. It is equivalent to using a single quadratic penalty at midpoint $\mathbf{y} = (\mathbf{y}_l + \mathbf{y}_u)/2$, with rate $2 \alpha$ vs. the usual $\alpha$. For this `interval deviance', there is clearly not a unique minimizer $\mathbf{x}$. This is not a problem if used as a penalty $\alpha \delta_2$ as proposed above instead of the base measure $\delta_1$. For our example, we can stack
$\delta_2(\mathbf{A}_2 \mathbf{x}_1|\mathbf{t}_l,\mathbf{Q}_2)$  and $\delta_2(\mathbf{A}_2 \mathbf{x}_1|\mathbf{t}_u,\mathbf{Q}_2)$, doubling the number of penalty terms (See Section \ref{sec:penalty}).
The analogy would be to have a flat ridge in the likelihood $\delta_2$, but have identifiability of parameters maintained due to the prior distribution $\delta_1$. In this case, a unique maximizer of the likelihood would not exist, but a unique maximizer of the posterior distribution would exist.

In addition to using the deviances mentioned, we can also partition $\mathbf{x}'=[\mathbf{x}'_1,\mathbf{x}'_2]$ and apply distinct deviances to each partition (for example logistic to $\mathbf{x}_1$ and quadratic to $\mathbf{x}_2$). The adjustment vector $\boldsymbol{\eta}$ is applied element-wise to the unconstrained value $\mathbf{y}$ to obtain $\mathbf{x}$. \emph{This does not mean that each element of the $\boldsymbol{\eta}$ adjustment vector is independent of each other or of $\mathbf{y}$}. It means that the \emph{final} step of solving for $\mathbf{x}$ is an element-wise operation, so we can compartmentalize the process to create “mixed” deviances. In other words, we can use distinct deviance families $\delta_1 (\mathbf{x}_1)$ and $\delta_2 (\mathbf{x}_2)$ but treat them in abstract for computation as $\delta (\mathbf{x})$. The real benefit here is the computational compartmentalization, where the first partition is the weights and the second are the controls (linear combinations of weights). The mixed deviance formulation allows us to implement different families of deviance measures for these two pieces. For example, we can now use a logistic deviance on the weights $\delta_1$  and a quadratic (or absolute difference) penalty on the controls $\delta_2$.

\subsection{Solution Equations and Solvers}\label{sec:solver}
The four smooth deviances each have the following properties: The derivatives 
$\mathbf{\partial_{x} \delta(x|y,Q)} = \mathbf{Q} \delta^{(1)}(\mathbf{x|y})$
 can be decomposed into a scaling matrix ($\mathbf{Q}$ or $\left< \mathbf{q} \right>$) times a vector of element-wise functions on $\mathbf{x}$ and $\mathbf{y}$. In particular, we define $\boldsymbol{\eta}=\delta^{(1)}(\mathbf{x|y})$. Then we have $\mathbf{x} = h[\boldsymbol{\eta}|\mathbf{y}]$ where $h$ is the element-wise inverse of $\delta^{(1)}(\mathbf{x|y})$. Next we consider the simplified optimization: 
\begin{equation}
\label{eq:opsimp}
\min_\mathbf{x} \left\{ \delta(\mathbf{x|y,Q}) \right\},  \text{ s.t. }  A \mathbf{x} = \mathbf{t}
\end{equation}
Using the method of Lagrange Multipliers \citep[see, for example][]{Stewart11}, a solution $\mathbf{x}$ will satisfy the condition that the derivatives of the objective function are parallel (equal up to a set of scaling factors $\boldsymbol{\lambda}$ to the derivatives of the constraint function: 
$\mathbf{Q} \delta^{(1)}(\mathbf{x|y}) = \mathbf{A}'\boldsymbol{\lambda}$
then
$\delta^{(1)}(\mathbf{x|y}) = \mathbf{Q}^{\minus 1} \mathbf{A}' \boldsymbol{\lambda}$.
Then applying $h[.]$ and $\mathbf{A}$ to each side, we arrive at solution equations for $\mathbf{x}$ and $\boldsymbol{\lambda}$:
\begin{equation}
\label{eq:est}
\begin{array}{r@{=}l}
\mathbf{x} & h[\mathbf{Q}^{\minus 1} \mathbf{A}' \boldsymbol{\lambda}]\\
\mathbf{t} & \mathbf{A} h[\mathbf{Q}^{\minus 1} \mathbf{A}' \boldsymbol{\lambda}]
\end{array}
\end{equation}
For the quadratic deviance, $h[\boldsymbol{\eta}] = \mathbf{y} + \boldsymbol{\eta}$, so the solution is closed form
\begin{equation}
\label{eq:quad.sol}
\mathbf{x} = \mathbf{y} + \mathbf{Q}^{\minus 1}\mathbf{A}'( \mathbf{A}\mathbf{Q}^{\minus 1}\mathbf{A}')^{\minus 1}(\mathbf{t} - \mathbf{Ay}).
\end{equation}
Otherwise, a solution can be found via iterative methods. From the set of equations (\ref{eq:est}), it can be shown that the solution $\mathbf{x}$ is invariant to full rank rotations of the rows of $\mathbf{A}$. For example, changing the reference levels for categorical variables within $\mathbf{A}$ will lead to identical results, as one might expect. It also can be shown that the solution $\mathbf{x}$ is invariant to multiplication by a finite scalar. Thus differential scaling by a vector is needed to produce alternative values for $\mathbf{x}$. There is an equivalence between replacing rows and columns in $\mathbf{Q}$ with zeros and adding equality constraints of the form $x_i = y_i$ to the $\mathbf{A}$ matrix \citep{Williams13}, which can also be used to produce modified solutions $\mathbf{x}$ (See Section \ref{sec:penalty}).

To solve for $\mathbf{x}$ in (\ref{eq:est}), we can apply Newton's Method to solve for $\boldsymbol{\lambda}$ \citep{Stewart11}. After convergence, we then evaluate $\mathbf{x} = h[\boldsymbol{\eta}^{j}]$ directly. For the four smooth deviances, the Newton step will be:
\begin{equation}
\label{eq:Newt}
\boldsymbol{\lambda}^{j+1} = \boldsymbol{\lambda}^{j} + \left[ \mathbf{A} \mathbf{Q}^{\minus 1}
	\left< h^{(1)} [\boldsymbol{\eta}^{j}] \right>
	\mathbf{A}' \right]^{\minus 1}
	 \times \left( \mathbf{t} - \mathbf{A} h[\boldsymbol{\eta}^{j}] \right) 
\end{equation}
with $\boldsymbol{\eta}^{j} = \mathbf{Q}^{\minus 1} \mathbf{A}' \boldsymbol{\lambda}^{j}$ and
$\partial_{\boldsymbol{\eta}} h(\boldsymbol{\eta}) = \left< h^{(1)}[\boldsymbol{\eta}]\right>$ is diagonal.
This is a direct application of Newton's method and is essentially the same as in \citet{Deville92}, with the matrix operations expressed more explicitly. It can also be expanded to a two-level method when constraints are non-linear \citep{Williams13}. Functions $h$ and $h^{(1)}$ for each of the smooth deviances are provided in Table \ref{tab:h}, which extends Table 1 in \citet{Deville92} by providing $h^{(1)}$ needed for computation and including explicit expressions for the logistic deviance in terms of the distribution functions.

\begin{table}
\centering
\caption{Mappings $\mathbf{x} = h[\boldsymbol{\eta}|\mathbf{y}]$ and derivatives $h^{(1)}[\boldsymbol{\eta}|\mathbf{y}]$ for four smooth deviances.}
\begin{tabular}{|l|c|c|}
\hline
{\bf Deviance} & $h[\boldsymbol{\eta}|\mathbf{y}]$ & $h^{(1)}[\boldsymbol{\eta}|\mathbf{y}]$\\ \hline
Quadratic & $ \mathbf{y} + \boldsymbol{\eta}$ & $\mathbf{1}$ \\ \hline
Poisson  & $ \mathbf{y}[\mathbf{1}- \boldsymbol{\eta}]^{\minus 1}$ & $ \mathbf{y}[\mathbf{1}- \boldsymbol{\eta}]^{\minus 2}$\\ \hline
Discrimination & $ \mathbf{y} \exp[\boldsymbol{\eta}]$ & $ \mathbf{y} \exp[\boldsymbol{\eta}]$ \\ \hline
Logistic & $\Phi_{log}\left[\boldsymbol{\eta} + \boldsymbol{\mu}  \right](\mathbf{b}_{u} - \mathbf{b}_{l}) + \mathbf{b}_{l}$ &
	$\phi_{log}\left[\boldsymbol{\eta} + \boldsymbol{\mu}  \right](\mathbf{b}_{u} - \mathbf{b}_{l})$\\
	&\multicolumn{2}{c|}{$\boldsymbol{\mu} =  \Phi^{\minus 1}_{log}\left[(\mathbf{y} - \mathbf{b}_{l})/(\mathbf{b}_u - \mathbf{b}_l)\right]$} \\ \hline
\end{tabular}
\label{tab:h}
\end{table}

Range restrictions of the form $\mathbf{b}_{l} \le \mathbf{x} \le \mathbf{b}_{u}$ can be incorporated in at least two ways. 
(i) We could choose the quadratic deviance and use quadratic programming approaches. The general quadratic program solves the following:
\begin{equation}
\label{eq:qp}
\min_\mathbf{x} \left\{\mathbf{x'Qx/2+x'c } \right\},  \text{ s.t. }  \mathbf{A}_{1} \mathbf{x} = \mathbf{t} \text{ and } \mathbf{A}_{2} \mathbf{x} \ge \mathbf{b}
\end{equation}
In our case $\mathbf{c}= -\mathbf{Qy}$ would correspond to the quadratic deviance. One major drawback for the application of weight adjustment is that inequalities for each $x_i$ need to be added as rows of $\mathbf{A}_{2}$. 
For more technical details and a variety of methods to solve for $\mathbf{x}$, see Chapter 16 of \citet{Nocedal06}. 

(ii) Alternatively, we can choose the logistic deviance which automatically enforces $\mathbf{b}_{l} \le \mathbf{x} \le \mathbf{b}_{u}$. In contrast to quadratic programming, it does not increase the number of equations needed. We recommend Newton’s method for the applications discussed here due to its speed and flexibility and the availability of good starting points (Section \ref{sec:guide}). Modifications to the Newton method, such as the line search or trust region methods, may improve stability and convergence at the cost of reduced speed and added complexity \citep{Nocedal06}. The techniques in the remainder of this section manipulate equation systems \ref{eq:opsimp} and \ref{eq:est} and therefore apply to both the Newton method and quadratic programming methods.

\subsection{Augmentation, Scaling, and Rescaling}\label{sec:penalty}
A key insight from system \ref{eq:optim} is that all three pieces of the optimization (deviance, penalty, and equality constraints) are related. We can now reparameterize the original objective \ref{eq:optim} as a much simpler form \ref{eq:opsimp} which does not have explicit penalty terms (note labels $\mathbf{x}_1$ and $\mathbf{x}_2$):

\begin{equation}\label{eq:aug}
\begin{array}{rll}
\min_{\mathbf{x}_{1}}  \left\{ \delta_1(\mathbf{x}_{1}|\mathbf{y}_{1},\mathbf{Q}_{1}) \right.
 &+ \ \left. \alpha \delta_2(\mathbf{A}_{2} \mathbf{x}_{1}|\mathbf{t}_{2}, \mathbf{Q}_{2}) \right\},  &  
	\text{ s.t. }  \mathbf{A}_{1} \mathbf{x}_{1} = \mathbf{t}_{1},\\
&\equiv \min_\mathbf{x}  \left\{ \delta(\mathbf{x|d,Q}) \right\}  & \text{ s.t. }  \mathbf{A x} = \mathbf{t}.
\end{array}
\end{equation}
We first parameterize $\mathbf{x}_{2} = \mathbf{A}_{2} \mathbf{x}_{1}$ centered at $\mathbf{y}_{2} = \mathbf{t}_{2}$. We then stack $\mathbf{x}' = [\mathbf{x}_{1}', \mathbf{x}_{2}']$ and $\mathbf{y}' = [\mathbf{y}_{1}', \mathbf{y}_{2}']$. Next we define the block-diagonal scaling matrix $\mathbf{Q} = \left< \mathbf{Q_1}, \alpha \mathbf{Q}_{2}\right>$. 
We incorporate the constraints $\mathbf{A}_{1} \mathbf{x}_{1}=\mathbf{t}_{1}$ and $\mathbf{A}_{2} \mathbf{x}_{1} - \mathbf{x}_{2} =\mathbf{0}$ with constraint matrix $\mathbf{A} = 
	\left[\begin{array}{cc}
		\mathbf{A}_{1} &\mathbf{0} \\
		 \mathbf{A}_{2} & \minus \mathbf{I}
	\end{array} \right]$
, targets $\mathbf{t}' = [\mathbf{t}_{1}', \mathbf{0}']$, and multipliers $\boldsymbol{\lambda}' = [\boldsymbol{\lambda}_{1}', \boldsymbol{\lambda}_{2}']$. 
Here we note that if $\mathbf{A}_{1}$ is full (row) rank, then $\mathbf{A}$ will also be.
This is the basis of the ridge stabilization method for the quadratic \citep{Rao97, Montanari09} and other deviances \citep{Bocci08} and the related mixed model formulation \citep{Park09, Breidt17}. Ridge stabilization has $\mathbf{t}_1$ empty with all controls in $\mathbf{t}_2$. The mixed model formulation has `fixed' controls in $\mathbf{t}_1$ and `random' controls in $\mathbf{t}_2$. By distributing the penalty term into the augmented base measure and the augmented equality constraints, the objective function is now parameterized as the simplified second line of equation \ref{eq:aug} and the optimization methods in Section \ref{sec:solver} can be directly applied. Like other augmented or latent-variable methods, such as the Expectation Maximization algorithms \citep{Dempster77} and the Gibbs sampler for probit regression \citep{Albert93}, this approach trades an increase in dimension for simplicity and more direct computation. If we require additional range restrictions we have at least two ways to generate solutions for system \ref{eq:opsimp}; the Newton method for logistic deviance and quadratic programming for the quadratic deviance. These would also provide reasonable starting points when using the other deviances both in terms of initial values for $\mathbf{x}$ and in identification of active constraints to include in $\mathbf{t}_1$ vs. $\mathbf{t}_2$. 
The augmented form \ref{eq:aug} also allows for direct manipulation of the scaling factors $\mathbf{Q}_2$ corresponding to targets.

Although typically thought of as weights (e.g. weighted least squares), we refer to $\mathbf{Q}$ and its diagonal version $\left< \mathbf{q} \right>$ as scaling factors to avoid confusion when working with sampling weights ($\mathbf{y = d}$ and $\mathbf{x = w}$).
There are often default choices for $\mathbf{Q}$ such as the identity matrix or the inverse of a variance-covariance matrix. In addition, we can manipulate and rescale portions of $\mathbf{Q}$ to introduce additional functionality into our original system $\ref{eq:est}$.
To see how the scaling matrix $\mathbf{Q}$ relates to other system components, we first partition $\mathbf{x}' = [\mathbf{x}'_{A}, \mathbf{x}'_{C}]$. 
The corresponding scale matrix is then 
$\mathbf{Q} = \left[\begin{array}{cc}
	\mathbf{Q}_{A} & \mathbf{Q}_{B} \\
	\mathbf{Q}_{B}' & \mathbf{Q}_{C}
	\end{array}\right]$.
Next we define 
$\mathbf{Q}_{0} = \left[\begin{array}{cc}
	\mathbf{Q}_{A} & \mathbf{0}\\
	\mathbf{0}' & \mathbf{0}
	\end{array}\right]$
and
$\mathbf{Q}_{\alpha} = \left[\begin{array}{cc}
	\mathbf{Q}_{A} & \alpha^\rho \mathbf{Q}_{B} \\
	\alpha^\rho \mathbf{Q}_{B}' & \alpha \mathbf{Q}_{C}
	\end{array}\right]$  for $\alpha > 0$ and $\rho < 1/2$.
Based on these alternative $\mathbf{Q}$ matrices, we have two very useful properties: (i) 
Using the Moore-Penrose generalized inverse 
$\mathbf{Q}_{0}^{\minus} = \left[\begin{array}{cc}
	\mathbf{Q}_{A}^{\minus 1} & \mathbf{0}\\
	\mathbf{0}' & \mathbf{0}
	\end{array}\right]$
in place of $\mathbf{Q}^{\minus 1}$ in the solution equations (\ref{eq:est}) is \emph{equivalent} to adding the equality constraint $\mathbf{x}_{C} = \mathbf{y}_{C}$ \citep{Williams13}.
(ii) Replacing $\mathbf{Q}^{\minus 1}$ with $\mathbf{Q}_{\alpha}^{\minus 1}$ in system \ref{eq:est} penalizes $\mathbf{x}_{C} \ne \mathbf{y}_{C}$ by a factor $\alpha$. As $\alpha \rightarrow \infty$, $\mathbf{Q}_{\alpha}^{\minus 1} \rightarrow \mathbf{Q}_{0}^{\minus}$ (See Appendix \ref{app:limits}). Thus the limit of penalized equations is equality constrained equations.

As mentioned in Section \ref{sec:absdev}, the non-smooth deviances may be considered extensions of the smooth deviances. Since the absolute difference deviance is not differentiable at $\mathbf{x=y}$, it does not fit directly into the simple derivative-based approach \ref{eq:est}. However, we can solve this system iteratively by using a smooth deviance and rescaling $\mathbf{Q}$. In particular, we can use the quadratic deviance 
$\delta(\mathbf{x^{i + 1}|y,Q}) = 
		(\mathbf{x}^{i + 1}-\mathbf{y})'\left<\boldsymbol{\kappa}^i \right>^{\minus 1}
		\left<\mathbf{q} \right> (\mathbf{x}^{i + 1}-\mathbf{y})$
where we update $\boldsymbol{\kappa}^i =|\mathbf{x}^{i}-\mathbf{y}|$ and solve for $\mathbf{x}^{i + 1}$ until convergence. This can be justified analogously to the implementation of penalized least squares as local quadratic approximations \citep{Fan01} or more generally as minimizing a local bounding function \citep{Lange00, Hunter04}. Here the absolute difference deviance is iteratively bounded locally by a quadratic deviance $\delta(\mathbf{x}^{i}|\mathbf{y,Q})$, which is then minimized and replaced by an updated bounding deviance $\delta(\mathbf{x}^{i + 1}|\mathbf{y,Q})$. Although a value of $\boldsymbol{\kappa}^{0}=\mathbf{1}$ is a good starting point, if we repeat the process for slightly different values of $\mathbf{Q}=\alpha_{k} \left< \mathbf{q} \right>$,  using values of $\mathbf{x}$ (and $\boldsymbol{\lambda}$) from similar values of $\alpha_{k-1}$, then convergence may occur more rapidly and iterations would be more stable for large $\alpha$ (See Section \ref{sec:guide}).

\subsection{Synthesis}\label{sec:synth}
Combining the conceptual and computational components of the previous sections, we propose a general strategy which caters to exploratory investigations but can be adjusted for operational production.

\subsubsection*{Identify Conflicting Targets and Plausible Ranges for Weights}
A common starting point for a new weight calibration process is to first use the quadratic deviance and all targets of interest without any range restrictions. \citep{Chen02} suggests using the original weights and initial solutions along with a bisection type approach to relax targets while trying to achieve more range restrictions. However, there may be problems with even getting a solution $(\alpha = \infty)$ if too many targets are specified or corresponding cell counts are small. We can construct a more informative path from $\mathbf{y}=\mathbf{d}$ to $\mathbf{x} = \mathbf{w}$, and explore trade-offs between how close a target needs to be versus how many boundaries are crossed and the increase in dispersion of the weights for a given $\alpha$.

In particular, we start with a quadratic deviance $\delta_1$  and quadratic penalty $\delta_2$. Using the augmentation and scaling techniques of Section \ref{sec:penalty}, we set $\mathbf{A}=[\mathbf{A}_{1}, \minus \mathbf{I}]$ with $\mathbf{A}_{2}=\mathbf{A}_{1}$, $\mathbf{y}_{2}=\mathbf{t}_{1}$ and $\mathbf{t}=\mathbf{0}$. Thus all targets $\mathbf{t}_{1}$ are set to penalties. If we wanted to enforce some constraints exactly, such as the sum of the weights, we could use the more general form of $\mathbf{A}$. As a default, we might choose $\mathbf{Q}_{1}  \propto  \left< \mathbf{y}_{1} \right>^{\minus 1}$ and $\mathbf{Q}_{2} \propto \mathbf{I}$ which puts the weights on a relative scale and controls on an absolute scale. We select $\alpha$ from a grid of $2^k$ with $k \in \{-m,\ldots,m\}$. We then set $\mathbf{Q} = \left< \mathbf{Q}_{1}, \alpha \mathbf{Q}_{2} \right>$ and solve \ref{eq:aug} for each $\alpha$. Evaluating $\mathbf{A}_{1} \mathbf{x}(\alpha)$ scaled by $\mathbf{t}_1$ and $\mathbf{x}(\alpha)$ vs. $\alpha$ provides useful representations of how weights change and targets are approached as the penalty increases. They indicate when targets cannot be met and when boundaries for weights are crossed. Evaluating weight dispersion vs. $\alpha$ also provides an indication of potential increase in variances of estimates. When not all targets are met, iteratively rescaling $\mathbf{Q}_{2}$ by $\mathbf{\kappa}^{i}=|\mathbf{x}_{2}^{i}-\mathbf{t}_{1} |$ will provide an alternative path in which a subset of these targets can be met at the expense of pushing others farther away.

\subsubsection*{Enforce Range Restrictions}
After using the quadratic deviance to explore the relationship between the targets and weights, we should have a better idea of which weight ranges are plausible. As noted in Section \ref{sec:solver} we can use either the logistic deviance or quadratic programming. For quadratic programming, we use the augmented equations with both the deviance and the penalty as quadratic. We use the same $\mathbf{A}$, $\mathbf{y}$, $\mathbf{t}$, and $\mathbf{Q}_{2}$ inputs. For the Newton method, we use a mixed deviance by combining a logistic deviance for the weights with a quadratic penalty for the targets and choose $\mathbf{Q}_{1} \propto \mathbf{I}$. 

\subsubsection*{Replace point targets with interval targets}
In addition to range restrictions, interval targets of the form $\mathbf{t}_{l} \le \mathbf{A}_{2} \mathbf{x}  \le \mathbf{t}_{u}$ may be applicable. For example, the estimate of farm land in a county must be between 0 and the total size of the county. Alternatively, controls for small domains often have non-negligible measurement error. Enforcing these inequalities exactly can be accommodated directly with quadratic programming. However, this approach may lead to an empty feasible region. Instead, we can enforce a set of interval penalties. As described in Section \ref{sec:dev}, an absolute difference penalty on the interval $[\mathbf{t}_{l},\mathbf{t}_{u}]$ can be enforced by applying an absolute difference penalty at each of the end-points with a common scaling factor for both. Analogous to the point version, this will pull non-compliant targets towards control bounds.

\section{NSDUH Post-stratification}\label{sec:app}
\begin{figure}
\centering
\includegraphics[width = 0.90\textwidth,
		page = 1,clip = true, trim = 0in 0.5in 0in 0.75in]{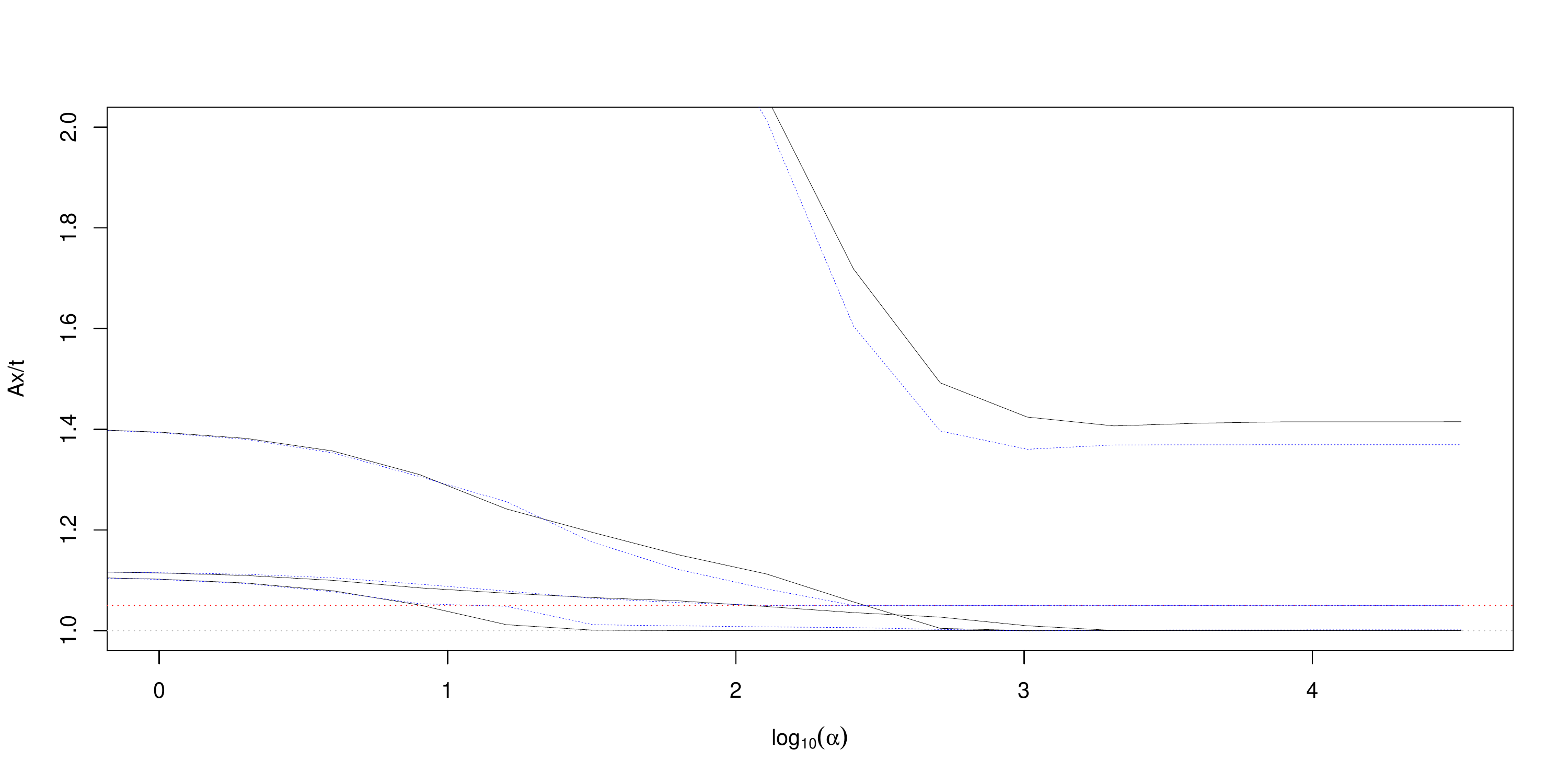}\\
\includegraphics[width = 0.90\textwidth,
		page = 1,clip = true, trim = 0in 0.5in 0in 0.75in]{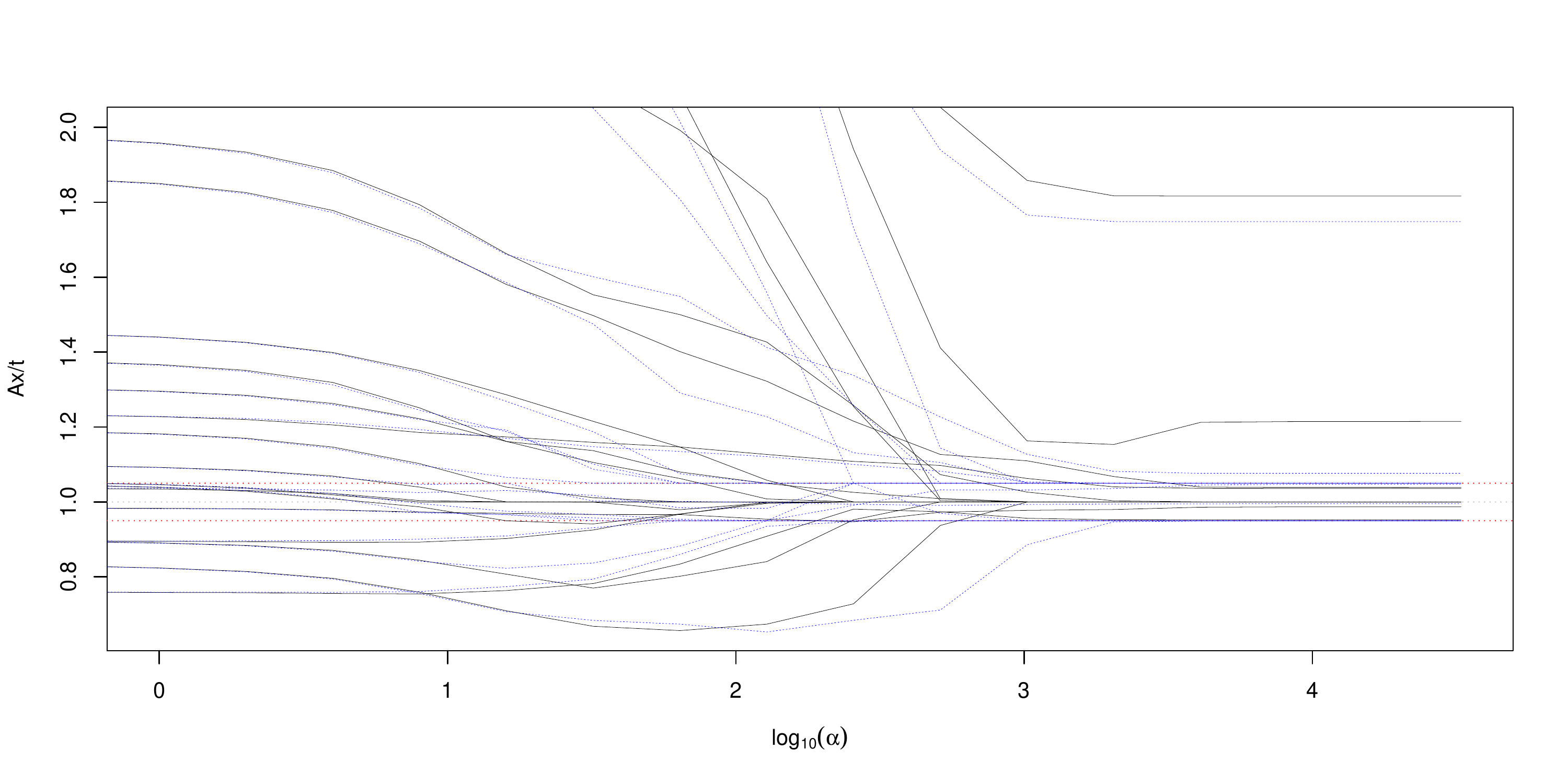}\\
\includegraphics[width = 0.90\textwidth,
		page = 1,clip = true, trim = 0in 0.25in 0in 0.75in]{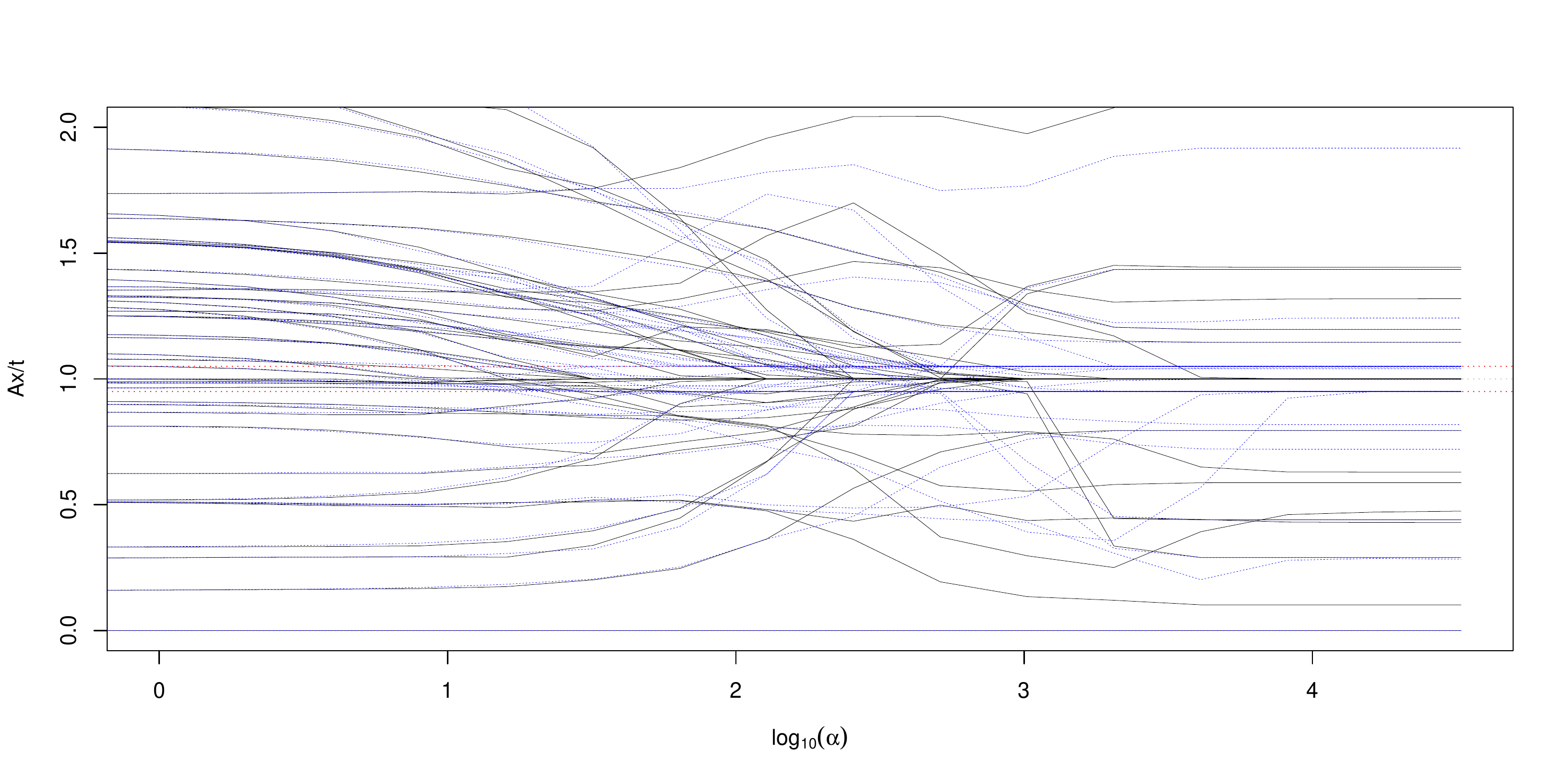}
\caption{Solution path for race controls $\mathbf{Ax}$ relative to target values $\mathbf{t}$ across increasing penalty $\alpha$ (left to right). Race (top), race by state (middle), race by age by state (bottom). Comparing LL1 (point controls, solid black) and LL1N (5\% interval controls, dashed blue). Exact achievement and achievement within 5\% tolerance (dotted grey and red lines).}
\label{fig:iterationLL1}
\end{figure}

We revisit the example from Section \ref{sec:motiv}, the 2014 NSDUH final post-stratification step for the New England division. Of the 267 potential controls, Table \ref{tab:analwt} shows that 196 were met, missing 71 controls. However, this count is too low, because some control categories were collapsed. In a strict sense, 105 controls were missed. Table \ref{tab:compare} shows that virtually all these missed controls involved Race, either as a 5-level or 3-level recode. As mentioned in Section \ref{sec:motiv}, the rank of the predictor matrix $\mathbf{A}$ is estimated as 257, so we expect that a solution that misses more than 10 but fewer than 105 controls is possible. The current NSDUH analysis weights (AW) are obtained by fitting a logistic deviance with specified upper and lower bounds based on the distribution of the weights from previous adjustment steps and the need to control for extreme weights. For simplicity and comparability, we use the same bounds for the alternatives presented here. Since this is an annual survey rather than a new data collection, the first step in Section \ref{sec:synth}, identifying conflicting targets (Race) and plausible ranges for weights, has already been done for us. We compare several alternatives which all enforce the provided range restrictions for every weight:
\begin{itemize}
\item	LL1: Logistic deviance base measure for the weights with absolute difference penalty for the targets.
\item	LL1N: Same as LL1, with interval penalties used for all Race factors and interactions. Intervals are $\pm 5$\% of the Census control total.
\item	LL2: Logistic deviance base measure for the weights with quadratic penalty for the targets
\item	QPL2: Quadratic deviance base measure for the weights with quadratic penalty for the targets with equal scaling and additional restrictions for weights enforced by augmenting the constraint matrix and using quadratic programming methods.
\end{itemize}
For simplicity, comparisons in Table \ref{tab:compare} are based on the number of targets missed, defined in terms of being (i) within rounding (nearest 1,000) of the Census total, or (ii) within 5\% of the target. As discussed in Section \ref{sec:intro}, there are other ways to evaluate weights and the impact on estimation; however, the purpose of this example is identify the space of achievable models (controls) rather than to compare the performance of models, which is a natural next step in an analysis. The LL1 appears to achieve many more controls than AW, missing only 53 in total. If we also include targets that are within 5\% of their Census controls, than all but 34 constraints are met. LL1N provides a small improvement over LL1 leading to 30 constraints being outside of 5\% of the target. Figure \ref{fig:iterationLL1} displays the solution paths for the constraints related to race for both LL1 and LL1N. Both have similar results; however, where race controls using LL1N are collapsed exactly to their targets at 100\% (when achieved), race controls using LL1N are less restricted, often collapsing only to the boundary values of 95\% and 105\%. This allows for some additional flexibility leading to some missed controls moving a little closer to their target values. Figure \ref{fig:iterationLL1} also reveals that some race by age by state controls are fixed at 0. These controls have no observations in the sample, but have population level targets that are non-zero. They represent unachievable targets (rank deficient $\mathbf{A}$) which could be removed. However, it is comforting that the procedure is robust to a certain amount of these inconsistencies and can easily identify them graphically.

Both the LL2 and QPL2 methods lead to very similar results with the same number of targets being missed per variable. They do much worse than the L1 methods in terms of exactly meeting targets, missing 142, but do well keeping targets within 5\% of the control, with only 44 falling outside this range. The main advantage of the L2 methods are that they are simpler to implement. However, the L1 methods are just modified L2 methods (as in Section \ref{sec:approach}), so they are also relatively simple. For a relative speed comparison, the LL1 and LL1N methods each took under 30s to run on a personal computer using sparse matrices with the `Matrix' library \citep{RMatrix} in R \citep[version x64 3.4.2,][]{R17} and a single processor thread using values $\alpha=2^k$ with $k \in \{-14,\ldots,15\}$. The added RAM used was negligible to the overhead from loading R and the data set. The LL2 method ran in about 4s over the same $\alpha$ values. The QPL2 method took around 4 \emph{hours} to run using `solve.QP' from the `quadprog' package \citep{RQP}, requiring an extra 1 to 2GB of RAM depending on the $\alpha$  iteration. The QP method could only use dense matrices and required over 11,000 additional constraints, two for every weight.

We note that all the alternatives presented in Table \ref{tab:compare} have one level of the one-factor Race variable which missed its target. The final weights (AW) do hit this target, because the NSDUH models are fit in succession with the one-factor models, followed by the two- and three-factor. Hierarchical rules are enforced for AW, so models with higher level interactions are forced to include the same levels as the previous lower level models. In contrast, the alternatives presented here are direct simultaneous implementations. Forcing some constraints to be met exactly can be implemented by using distinct $\mathbf{A}_1$ and $\mathbf{A}_2$ matrices as presented in general in Section \ref{sec:approach}, rather than the special case of $\mathbf{A}_2=\mathbf{A}_1$ used here for simplicity of discussion.

\begin{table}
\begin{center}
\caption{Demographic variables for final post-stratification for New England for the 2014 NSDUH.
\emph{AW.} Final analysis weight. \emph{LL1.} Logistic deviance and absolute difference penalty \emph{LL1IN.} Same as LL1 with interval penalty for all Race variables ($\pm 5\%$) \emph{LL2.} Logistic deviance and quadratic penalty \emph{QPL2.} Quadratic programming with quadratic penalty. $^1$ Variables with weighted totals more than 1 unit (1,000’s) different from Census total. LL1N race categories are missed if weighted totals are more than 1 unit outside the ($\pm 5\%$) interval. $^2$ Variables with weighted totals more than $5\%$ different from Census totals.}
{\scriptsize
\begin{tabular}{|lr|*{5}{r}|*{5}{r}|}
  \hline
   &  & \multicolumn{5}{c|}{{\bf Missed}$^1$}  & \multicolumn{5}{c|}{{\bf Missed by} $\mathbf{> 5\%}^2$} \\
{\bf Variable} & {\bf Levels} & {\bf AW} & {\bf LL1} & {\bf LL1IN} & {\bf LL2} & {\bf QPL2} & {\bf AW} & {\bf LL1} & {\bf LL1IN} & {\bf LL2} & {\bf QPL2} \\ \hline
\emph{One-Factor Effects} & \emph{20} & \emph{0} & \emph{1} & \emph{1} & \emph{6} & \emph{6} & \emph{0} & \emph{1} & \emph{1} & \emph{1} & \emph{1}\\ \hline
Intercept & 1 & 0 & 0 & 0 & 0 & 0 & 0 & 0 & 0 & 0 & 0\\
State & 5 & 0 & 0 & 0 & 1 & 1 & 0 & 0 & 0 & 0 & 0\\
Quarter & 3 & 0 & 0 & 0 & 0 & 0 & 0 & 0 & 0 & 0 & 0\\
Age & 5 & 0 & 0 & 0 & 1 & 1 & 0 & 0 & 0 & 0 & 0\\
Race (5 levels) & 4 & 0 & 1 & 1 & 4 & 4 & 0 & 1 & 1 & 1 & 1\\
Gender & 1 & 0 & 0 & 0 & 0 & 0 & 0 & 0 & 0 & 0 & 0\\
Hispanicity & 1 & 0 & 0 & 0 & 0 & 0 & 0 & 0 & 0 & 0 & 0\\ \hline
\emph{Two-Factor Effects} & \emph{95} & \emph{11} & \emph{14} & \emph{8} & \emph{50} & \emph{50} & \emph{11} & \emph{6} & \emph{5} & \emph{12} & \emph{12}\\ \hline
Age x Race (3 levels) & 10 & 0 & 5 & 2 & 10 & 10 & 0 & 4 & 2 & 2 & 2\\
Age x Hispanicity & 5 & 0 & 0 & 0 & 1 & 1 & 0 & 0 & 0 & 0 & 0\\
Age x Gender & 5 & 0 & 0 & 0 & 1 & 1 & 0 & 0 & 0 & 0 & 0\\
Race (3 levels) x Hispanicity & 2 & 0 & 0 & 0 & 2 & 2 & 0 & 0 & 0 & 0 & 0\\
Race (3 levels) x Gender & 2 & 0 & 0 & 0 & 2 & 2 & 0 & 0 & 0 & 0 & 0\\
Hispanicity x Gender & 1 & 0 & 0 & 0 & 0 & 0 & 0 & 0 & 0 & 0 & 0\\
State x Quarter & 15 & 0 & 0 & 0 & 0 & 0 & 0 & 0 & 0 & 0 & 0\\
State x Age & 25 & 0 & 2 & 3 & 11 & 11 & 0 & 0 & 0 & 0 & 0\\
State x Race (5 levels) & 20 & 11 & 7 & 3 & 20 & 20 & 11 & 2 & 3 & 10 & 10\\
State x Hispanicity & 5 & 0 & 0 & 0 & 1 & 1 & 0 & 0 & 0 & 0 & 0\\
State x Gender & 5 & 0 & 0 & 0 & 2 & 2 & 0 & 0 & 0 & 0 & 0\\ \hline
\emph{Three-Factor Effects} & \emph{152} & \emph{94} & \emph{38} & \emph{29} & \emph{86} & \emph{86} & \emph{58} & \emph{27} & \emph{24} & \emph{44} & \emph{44}\\ \hline
Age x Race (3 levels) x Hispanicity & 10 & 4 & 5 & 2 & 10 & 10 & 2 & 4 & 2 & 2 & 2\\
Age x Race (3 levels) x Gender & 10 & 4 & 3 & 1 & 7 & 7 & 2 & 1 & 1 & 3 & 3\\
Age x Hispanicity x Gender & 5 & 0 & 0 & 0 & 1 & 1 & 0 & 0 & 0 & 0 & 0\\
Race (3 levels) x Hispanicity x Gender & 2 & 0 & 0 & 0 & 2 & 2 & 0 & 0 & 0 & 0 & 0\\
State x Age x Race (3 levels) & 50 & 50 & 20 & 18 & 39 & 39 & 48 & 20 & 18 & 31 & 31\\
State x Age x Hispanicity & 25 & 20 & 7 & 5 & 11 & 11 & 1 & 0 & 0 & 0 & 0\\
State x Age x Gender & 25 & 0 & 0 & 0 & 0 & 0 & 0 & 0 & 0 & 0 & 0\\
State x Race (3 levels) x Hispanicity & 10 & 8 & 3 & 2 & 10 & 10 & 1 & 2 & 2 & 5 & 5\\
State x Race (3 levels) x Gender & 10 & 6 & 0 & 1 & 4 & 4 & 4 & 0 & 1 & 3 & 3\\
State x Hispanicity x Gender & 5 & 2 & 0 & 0 & 2 & 2 & 0 & 0 & 0 & 0 & 0\\ \hline
\emph{TOTAL} & \emph{267} & \emph{105} & \emph{53} & \emph{38} & \emph{142} & \emph{142} & \emph{69} & \emph{34} & \emph{30} & \emph{57} & \emph{57} \\ \hline
\end{tabular}
\label{tab:compare}
}
\end{center}
\end{table}
  
\FloatBarrier

\section{Practical Guidance}\label{sec:guide}
From the previous section, we have seen that the Newton solver approach works well for a moderate sized data set (5,000+ records, and 250+ constraints). It also scales up well to larger data sets. For example a more complex version of LL1 and LLN ran on 100,000+ records with 200+ controls in about 15 min on a single processor thread with less than 1 GB of available RAM \citep{Lamas15}. To reiterate, the two main reasons for this scalability are (i) the use of the logistic deviance to automatically enforce (hundreds of) thousands of range inequalities and (ii) the direct incorporation of sparse matrix manipulations, particularly sparse matrix multiplication. Thus the effective sizes of the system of equations and the matrix operations are much smaller than their nominal sizes. Since many of these controls are for sub-domains based on categorical variables, the $\mathbf{A}$ matrix is relatively sparse. Using sparse matrix structures can save memory and speed up computation significantly. In principle, quadratic programming could also make use of this sparsity, but its incorporation is more complex. Additionally, the (hundreds of) thousands of range restrictions would greatly increase the dimension of the system of equations.

In some applications, such as coverage adjustments for a census \citep{Lamas15}, the required range restrictions are quite narrow. In practice, a quadratic penalty $\delta_2$ performs poorly in this setting. It attempts to balance all the controls at once and leads to weights that quickly (small $\alpha$) hit their range boundaries without much progress being made toward selecting individual controls. In contrast, the absolute difference penalty is able to discriminate between competing controls and can thus progress much farther (large $\alpha$) before a large enough fraction of weights hit their range boundaries to cause numerical instability.

For any  large scale system of equations, numerical stability is critical. Good starting values are especially important for Newton methods.  Thus we recommend using the solution $\mathbf{x}(\alpha_{k-1})$ and Lagrange multiplier $\boldsymbol{\lambda}(\alpha_{k-1})$ from the previous iteration with $\alpha_{k-1}=2^{k-1}$   as starting values for the next step $\alpha_k$. In the examples discussed here, a simple doubling of $\alpha$ seems sufficient; however, more sophisticated rules for increasing the penalty are available \citep{Nocedal06}. For simplicity, our stopping rule was fixed $(\alpha =2^m)$. In practice, we may keep increasing $\alpha$ until the procedure reaches numeric instability and fails. Alternatively, we could stop if our current solution $\mathbf{x}(\alpha)$ is a stationary point of the penalty function $\delta_2$ (within a tolerance). Definitions of stationary points for both quadratic and absolute difference functions are available \citep{Nocedal06} and might be extended to this setting. We might also consider informally assessing stationarity with plots of parameters $\boldsymbol{\lambda}$, $\mathbf{x}$, and $\mathbf{Ax}$ vs. $\alpha$, stopping when slopes are close to 0 for large $\alpha$. For example, Figure \ref{fig:iterationLL1} shows that controls have stabilized for the last several iterations of $\alpha$.

Another useful tool to increase stability is to set some minimum thresholds below which weights are exactly equal to one of their boundaries and targets exactly equal to their controls. Setting weights and targets can be done in several ways \citep{Williams13, Pizzinga10} including setting the corresponding scaling factors to 0, augmenting the equality constraints, and reducing the system of equations. For large data sets, there is a trade-off. Creating new matrices in memory and copying values from the old at each $\alpha$ iteration may not be trivial in terms of computation time, even if the resulting system is smaller and more stable. Likely a hybrid approach that sets a criterion for when to trim down the system would be needed.

\section{Extensions and Future Work}\label{sec:ext}
While the above discussion and examples focus on a single step adjustment, multiple steps of sample weight adjustment are common (e.g. NSDUH). One alternative to consider is to simultaneously adjust for all of these components rather than using a series of adjustments \citep{Slud13}. While the work presented here focused on sampling weight adjustments, many of the developments and ideas overlap with the benchmarking of model-based estimates. For example, multiple levels of benchmarking for model-based estimates \citep{Ghosh13} may be represented by multiple penalty terms added to the objective function in equation \ref{eq:optim}. Furthermore, recasting equality constraints as penalties can also be viewed as dealing with external calibration or benchmarking targets which are measured with error \citep{Bell12}. It is our hope the techniques presented here can also be used to enhance benchmarking methods.

The Lagrange system \ref{eq:est} extends naturally to include differentiable non-linear constraints such as ratios \citep{Williams13} and variances \citep{Ghosh92, Datta11}, sometimes with closed form solutions. The Newton method from Section \ref{sec:solver} can be extended by iteratively updating $\mathbf{x}=h[\boldsymbol{\eta}(\mathbf{x})]$ which is now a function of $\mathbf{x}$. Alternatively, we can iteratively apply linear constraints which approximate non-linear ones \citep[see][ch 12]{Knottnerus03}. One example would be to model non-response propensities directly as $\mathbf{x} = \boldsymbol{\rho} \in [0,1]$  with non-linear constraints $g(\mathbf{x})= \mathbf{A}\mathbf{x}^{\minus 1}=\mathbf{t}$, which could be solved with a two-level Newton method iterating both $\mathbf{x}$ and $\boldsymbol{\lambda}$ \citep{Williams13}. These non-linear methods seem to work well for simple functions such as variances and ratios, but there is much room for investigation and improvement.

The methods developed here and the extensions mentioned above fit into the general framework of constrained optimization \citep{Nocedal06}. Although the deviances considered here are convex and the equality constraints are linear, a formal exploration of the Karush-Kuhn-Tucker (KKT) stationarity conditions may be fruitful, especially if we expand to some non-linear functions like variances and ratios. It may be straight forward to show that for a given $\alpha$, the stationary point $\mathbf{x}(\alpha)$ is a local (or even global) solution for the augmented system \ref{eq:aug}, but it is less clear how to interpret these conditions when evaluating the limit of this approach $(\alpha \rightarrow \infty)$, since the original fully constrained and restricted system typically has no solution.

Our approach implicitly assumes that the solution path $\mathbf{x}(\alpha)$ is a continuous function of $\alpha$. For the special case when all constraints can be achieved, it is clearly true for quadratic deviances and penalties \ref{eq:quad.sol} and could be extended to other deviances. However, it is unclear what happens in the case where no solution exists for the original problem. It is uncertain that a unique, continuous solution path $\mathbf{x}(\alpha)$ is guaranteed to exist. But it seems plausible that the approximate path generated by the procedure we develop here can still be useful as a partial or suboptimal solution. 

There are connections between these issues and path algorithms for constrained regression \citep{Tibshirani11, Zhou13} and data-cloning methods for evaluating parameter identifiability \citep{Lele07, Lele10}. In addition, the similarity in results for using the logistic deviance and quadratic programming also suggest that large quadratic programming systems might be efficiently approximated with the use of a logistic deviance for simple inequalities and interval penalties for more complex inequalities. More work is needed at the intersection of these areas. 

\section{Conclusions}\label{sec:conc}
Even with the need for more research, the framework and supporting methods developed in this paper show extensive applicability. This framework provides the flexibility to compromise in a systematic but customizable way including the
\begin{itemize}
\item	Use of all the popular deviance (distance) measures
\item	Use of range restrictions for weights 
\item	Use of point and interval constraints (controls)
\item	Ability to identify and target subsets of weights  and constraints which may be driving non-existence of a solution
\end{itemize}
As discussed in Section \ref{sec:guide}, the methods scale up well to realistic data requirements. 

While unique and optimal solutions are ideal, it is often desirable to consider alternatives that give useful solutions when unique ones are not available. This distinction is relevant when applying these algorithms to the regular production of official statistics. Having an efficient method that reduces dimensions and points the analyst towards a manageable number of alternatives should be competitive with the current methods in the field which often entail highly specific prioritization of targets or complex `guess and check' heuristics. 

As we note in the introduction, optimization for weight construction and adjustment is only one component of the process of making inferences from survey data. Therefore an efficient, automated process may be preferred to time-consuming iteratively-constructed customizations. Although the working models (deviances) described here are used extensively for survey weight adjustments, they are very simple and may not seem adequate for other statistical modeling applications. However, this simplicity allows us to explore the solution space and can later be replaced by more complex models once the insights from active constraints and restrictions are better understood.

\vskip 0.2in
\noindent%
{\it Acknowledgements:}
Special thanks to Lanting Dai and RTI International for assistance with the NSDUH files.
\vfill

\appendix

\section{Limiting results for penalized equations}\label{app:limits}
\begin{lemma}
\label{lem:Walpha}
Define
\[
\mathbf{Q}_\alpha = \left[
\begin{array}{cc}
\mathbf{Q}_A & \alpha^\rho \mathbf{Q}_B \\
\alpha^\rho \mathbf{Q}_{B}' & \alpha \mathbf{Q}_C
\end{array}
\right]
\]
where $\alpha > 1$, $\rho < 1/2$, and $\mathbf{Q}_A$ and $\mathbf{Q}_C$ are invertible.
Then
\[
	\lim_{\alpha \rightarrow \infty} \left\{ \mathbf{Q}_\alpha ^{\minus 1} \right\} = \mathbf{Q}_0^{\minus}
\]
where 
\[
\mathbf{Q}_0^{\minus} = \left[
\begin{array}{cc}
\mathbf{Q}_{A}^{\minus 1} & \mathbf{0} \\
\mathbf{0} & \mathbf{0}
\end{array}
\right].
\]
\end{lemma}
Note that 
$\lim_{\alpha \rightarrow \infty} \left\{ \mathbf{Q}_\alpha ^{\minus 1} \right\} \ne 
\left\{ \lim_{\alpha \rightarrow \infty} \mathbf{Q}_\alpha \right\}^{\minus 1}$
since the latter is not well-defined.

\begin{proof}
By the block inversion formula
\[
\mathbf{Q}_\alpha^{\minus 1} = \left[
\begin{array}{cc}
\mathbf{B}_{11} &  \mathbf{B}_{12}\\
\mathbf{B}_{12}' &  \mathbf{B}_{22}
\end{array}
\right]
\]
with
\[
\begin{array}{rl}
\mathbf{B}_{11} & = \ (\mathbf{Q}_{A}  - \alpha^{2\rho-1}\mathbf{Q}_{B} \mathbf{Q}_{C}^{\minus 1} \mathbf{Q}_{B}' )^{\minus 1}  \\
\mathbf{B}_{12} & = -\mathbf{B}_{11} \mathbf{Q}_{B} \mathbf{Q}_{C}^{\minus 1} \alpha^{\rho-1} \\
\mathbf{B}_{22} & = \alpha^{\minus 1}  \mathbf{Q}_{C}^{\minus 1} + 
	\alpha^{2(\rho-1)} \mathbf{Q}_{C}^{\minus 1} \mathbf{Q}_{B}' \mathbf{B}_{11} \mathbf{Q}_{B}\mathbf{Q}_{C}^{\minus 1}
\end{array}
\]
It follows that as $\alpha \rightarrow \infty$
\begin{itemize}
\item for $\rho < 1/2$,
 $\mathbf{B}_{11} \rightarrow \mathbf{Q}_{A}^{\minus 1}$,
 $\mathbf{B}_{12} \rightarrow \mathbf{0}$, and
 $\mathbf{B}_{22} \rightarrow \mathbf{0}$.
\item When $\rho = 1/2$, 
$\mathbf{B}_{11} \rightarrow (\mathbf{Q}_{A} - \mathbf{Q}_{B} \mathbf{Q}_{C}^{\minus 1} \mathbf{Q}_{B}')^{\minus 1}$,
 $\mathbf{B}_{12} \rightarrow \mathbf{0}$, and
 $\mathbf{B}_{22} \rightarrow \mathbf{0}$.
\item For $\rho >  1/2$,
 $\mathbf{B}_{11} \rightarrow \mathbf{0}$,
 $\mathbf{B}_{12} \rightarrow \mathbf{0}$, and
 $\mathbf{B}_{22} \rightarrow \mathbf{0}$.
\end{itemize}
\end{proof}

Using the augmentation approach from Section \ref{sec:penalty}, we create the following:
\begin{itemize}
	\item Let $\mathbf{y}' = [\mathbf{y}'_1, \mathbf{y}'_2 = \mathbf{t}'_1]$ and 
	$\mathbf{x}' = [\mathbf{x}'_1, \mathbf{x}'_2 = \mathbf{A}_{1} \mathbf{x}_1]$
	\item Set $\mathbf{Q}_\alpha = \left< \mathbf{Q}_1, \alpha \mathbf{Q}_2 \right>$ and $\alpha > 1$.
	\item Define the constraint $\mathbf{Ax} = \mathbf{A}_1 \mathbf{x}_1 - \mathbf{x}_2$, with target $\mathbf{t} = \mathbf{0}_k$. 
	Then the constraints $\mathbf{A} = [\mathbf{A}_1, \mathbf{\minus I}_k]$.
\end{itemize}
We now have the augmented equations:
\begin{equation}
\label{eq:aug_alpha}
\begin{array}{rl}
\mathbf{x} & = \ h[\mathbf{Q}_\alpha^{\minus 1} \mathbf{A}' \boldsymbol{\lambda}] \\
\mathbf{t} & = \ \mathbf{A}(h[\mathbf{Q}_\alpha^{\minus 1} \mathbf{A}' \boldsymbol{\lambda}]).
\end{array}
\end{equation}
As we increase the penalty $\alpha \rightarrow \infty$,
Lemma \ref{lem:Walpha} shows us that the solution equations (\ref{eq:aug_alpha}) converge to
\begin{equation}
\label{eq:aug_0}
\begin{array}{rl}
\mathbf{x} & = \ h[\mathbf{Q}_0^{\minus} \mathbf{A}' \boldsymbol{\lambda}] \\
\mathbf{t} & = \ \mathbf{A}(h[\mathbf{Q}_0^{\minus} \mathbf{A} \boldsymbol{\lambda}]).
\end{array}
\end{equation}
From the results in \cite{Williams13}, this is equivalent to setting $\mathbf{x}_2 = \mathbf{t}_1$. 
Since the constraint $\mathbf{A}_1 \mathbf{x}_1 = \mathbf{x}_2$ is also enforced, 
we have now induced the equality constraints $\mathbf{A}_1 \mathbf{x}_1 = \mathbf{t}_1$.

\vskip 0.2in
\bibliography{calibrationAug18}
\bibliographystyle{agsm}

\end{document}